\documentclass[iop]{emulateapj}
\usepackage{color,ulem}
\shorttitle{BH detection by Gaia}
\shortauthors{Yamaguchi et al.}

\usepackage{epsf}
\usepackage{epsfig}
\usepackage{amsmath}
\usepackage{amssymb}
\usepackage{graphicx}
\usepackage{booktabs}

\newcommand{\sm}{{\rm M}_{\odot}}
\newcommand{\gaia}{{\it Gaia}}
\newcommand{\mv}{M_{\rm V}}

\begin{document}

\title{Detecting Black Hole Binaries by Gaia}


\author{Masaki S. Yamaguchi}
\affil{Institute of Astronomy, The University of Tokyo, 2-21-2, Osawa, Mitaka, Tokyo, 181-0015, Japan}
\email{masaki@ioa.s.u-tokyo.ac.jp}

\author{Norita Kawanaka}
\affil{Department of Astronomy, Graduate School of Science, Kyoto University, Kitashirakawa Oiwake-cho, Sakyo-ku, Kyoto, 606-8502, Japan}

\author{Tomasz Bulik}
\affil{Astronomical Observatory, University of Warsaw, Al. Ujazdowskie 4, 00-478 Warsaw, Poland}

\author{Tsvi Piran}
\affil{Racah Institute for Physics, The Hebrew University, Jerusalem, 91904, Israel}

\begin{abstract}
We study the prospects of  the \gaia\ satellite to identify black hole (BH) binary systems by detecting the orbital motion of the companion stars.  Taking into account the initial mass function, mass transfer, common envelope phase, interstellar absorption and identifiability of the systems, we estimate the number of BH binaries that can be detected by \gaia\ and their distributions with respect to the BH mass. Considering  several models with different parameters we find that  $\sim$ 200-1,000 BH binaries could be detected by \gaia\ during its  $\sim 5~{\rm years}$ operation. The　 shape of the BH mass distribution function is affected strongly by the  zero-age main sequence (ZAMS) stellar mass  -  black hole mass relation. We show that once this distribution is established observationally we will be able to   constrain the currently unknown ZAMS mass - BH mass relation. 
\end{abstract}


\section{Introduction}
More than  $ 10^{8-9}$ stellar-mass black holes (BHs) are believed to reside in our Galaxy \citep{brownbethe94, timmes+96}. Today 59 X-ray binaries are considered to harbor a BH \citep{cor16}.   These are the only representatives of those numerous BHs that have been  observationally identified so far.   The masses of the BHs and the companion star in these binaries are estimated from optical and X-ray observations.  \cite{ozel+10} presented the Galactic black hole mass distribution which is based on the dynamically measured masses of 16 black holes in transient low-mass X-ray binaries.  The  observed BH mass distribution is quite narrowly centered around $7.8\pm 1.2~M_{\odot}$. BHs in the mass range of $2-5~M_{\odot}$ are absent in the Galaxy (at least as far as X-ray binaries are concerned).  \cite{farr+11} perform a Bayesian analysis using the observed black holes masses of 20 X-ray binaries and reach the similar results. 


LIGO has detected Binary BH mergers in distant galaxies \citep{LIGO16a, LIGO16b, LIGO17}. Somewhat surprisingly these BHs are significantly more massive than the observed Galactic X-ray binary BHs, although there were indications that the masses of BH binaries in gravitational waves should be higher than those in X-rays \citep{bel10,bul11}.  The corresponding masses range from $\sim 7.5M_\odot$ to $\sim 36M_\odot$. At present it is not clear if the difference in the observed mass distribution is due to a different origin of the progenitors, to differences in the initial mass function or simply due to the LIGO's sensitivity that is larger for more massive BHs.

In X-ray binaries, X-ray emission originates from a mass transfer from the companion star to the BH. Such a mass transfer is expected when the radius of a companion star is larger than the Roche lobe radius of the system.  However, if the orbital separation of a binary is too wide, there won't be a mass transfer and no X-ray emission.  The BH does not emit any electromagnetic radiation in such cases.  However, we can still discover such BH and estimate its mass if we are able to measure the orbital period and the semi-major axis with astrometry for its companion \citep{2017IAUS..324...41K,mashianloeb17,2017ApJ...850L..13B}. This idea is also mentioned in \gaia's white paper \citep{bar14}.

The astrometric satellite, {\it Gaia}, that was launched at the end of 2013 is an ideal tool to perform the needed observations.  {\it Gaia} can perform absolute astrometric measurement with a great precision on objects brighter than $G<20$ mag, where $G$-band covers wavelength between 0.3 and 1.0$\mu$m \citep{deb12}.
In case of a BH binary, if the companion is sufficiently bright, {\it Gaia} will detect its motion from which the existence of the BH can be inferred.  Our goal here is to estimate, following \cite{2017IAUS..324...41K} the expected number of BH binaries that can be detected by {\it Gaia} over its 5-year mission.  In this work we follow the binary formation and evolution taking into account the initial mass function, common envelope phase and mass transfer, and we estimate the total number of Galactic BH-main sequence star binaries without mass accretion in our Galaxy, as well as their distribution with respect to their masses and orbital separations. We then estimate their detectability and identifiability taking into account the interstellar absorption and obtain the number of such binaries detectable by {\it Gaia} during its operation ($\sim 5~{\rm years}$). We consider BH binaries that are bright enough to be detectable and their orbits are shorter than {\it Gaia}'s mission life time, yet they are far enough not to involve mass transfer so that they are not active X-ray sources.  Recently \cite{mashianloeb17} estimated the number of black hole binaries detectable by {\it Gaia} over its 5-year mission as nearly $2\times 10^5$.  However, they don't take into account the change of the orbital parameters due to  mass transfer from the  primary star (i.e., the black hole's progenitor) to the secondary or the common envelope phase.  In addition, they do not take into account the interstellar absorption. These effects could reduce significantly the number of detectable black hole binaries.  \cite{2017ApJ...850L..13B} also estimated the number of such black hole binaries using the binary population synthesis code {\sf COSMIC}, and predict that {\it Gaia} will be able to discover 3,800 to 12,000 black hole binaries.  However, they also don't take into account the interstellar absorption and so this value may be overestimated.  Moreover, it is not clear how their results depend on the relation between the initial stellar mass and the remnant black hole mass, which is closely related to the mechanism of core collapse supernovae.

The structure of this work is as follows. We begin in  \S \ref{sec:model} with a discussion of the initial conditions of BH progenitors and we derive the differential number of BH binaries detectable with \gaia\ per unit distance, BH mass, companion mass, and semi-major axis of BH binaries, where we take into account the binary evolution scenarios, focusing separately on low and high mass ratio binaries (\S \ref{sec:general}-\ref{sec:qsmall}). We then give the ranges of the companion mass and semi-major axis for integrating the differential number, considering the effect of interstellar extinction and the condition required for the detection of BH binaries with \gaia\ (\S \ref{sec:extinction} and \S \ref{sec:constrnts}). By integrating the differential number, we obtain the total number and the distribution of the detectable BH binaries for several models in \S \ref{sec:res}. In \S \ref{sec:discuss}, we discuss the possibility that the BH detected with \gaia\ constrains the relation between the ZAMS mass and BH mass. Finally, we conclude in \S \ref{sec:conclusion}.

\section{Model}
\label{sec:model}

We begin with a discussion of the evolution of  binaries, starting with 
a set of initial binary parameters: the two initial masses of the progenitors:   $\bar{M}_1$ and $\bar{M}_2$ (or alternatively the primary's mass and the mass ratio $q\equiv \bar{M}_1/\bar{M}_2$, with $q\le 1$) and the initial separation, $\bar A$. Hereafter barred quantities are initial ones and unbarred one are the present quantities e.g.  $M_{BH}$ and $M_2$ the current masses of the primary (that has turned into a BH) and the secondary, respectively.
In the following subsections, we show the total number of black hole-main sequence star binaries without mass accretion in our Galaxy, as well as their distribution with respect to their masses and orbital separation. We  consider two cases separately: a mass ratio close to unity ($q>0.3$) in \S \ref{sec:q1} and a small mass ratio ($q<0.3$) in \S \ref{sec:qsmall}.  The binaries in the former case would experience the mass transfer phase, while those in the latter case would experience the common envelope phase.  Although this division is too simplified, \cite{2012ApJ...746...22B, 2014ApJ...784...97B} adopted this assumption to derive the X-ray luminosity functions and binary-period distribution functions of high- and low-mass X-ray binaries, which are in agreement with the observations.  Thus we adopt this division in the current work.

\subsection{A general description}\label{sec:general}

We begin by considering  the distributions of the binary parameters at birth. 
For the primary star we use the Kroupa  \citep{kro01} initial mass function (IMF)  as the fiducial one: 
\begin{equation}
\Psi_{\rm K01}(\bar{M}_1)d\bar{M}_1\propto \left\{
\begin{array}{ll}
\bar{M}_1^{-1.3}d\bar{M}_1, & 0.08M_{\odot}\leq \bar{M}_1 < 0.5M_{\odot}, \\
\bar{M}_1^{-2.3}d\bar{M}_1, & 0.5M_{\odot}\leq \bar{M}_1 < 100M_{\odot}. \\
\end{array} \right.
\end{equation}
We also take another IMF given by \cite{kroupa+93} and \cite{kroupaweidner03} to investigate dependence of results on IMF:
\begin{equation}
\Psi_{\rm K93}(\bar{M}_1)d\bar{M}_1\propto \left\{
\begin{array}{ll}
\bar{M}_1^{-1.3}d\bar{M}_1, & 0.08M_{\odot}\leq \bar{M}_1 < 0.5M_{\odot}, \\
\bar{M}_1^{-2.2}d\bar{M}_1, & 0.5M_{\odot}\leq \bar{M}_1 < 1.0M_{\odot}, \\
\bar{M}_1^{-2.7}d\bar{M}_1, & 1.0M_{\odot}\leq \bar{M}_1 <100M_{\odot}. \\
\end{array} \right.
\end{equation}

For the secondary mass, we assume a flat mass ratio distribution as the fiducial case, \citep{kuiper35, kobulnickyfryer07}:
\begin{eqnarray}
\Phi(q)=\frac{1}{(1-\bar M_{\rm min}/\bar{M}_1)} q^{0}.
\end{eqnarray}
We also try the cases of the index of $q$, -1 and +1.
Here, we set the lower limit of $q$ as $\bar{M}_{\rm min}/\bar{M}_1$, where $\bar{M}_{\rm min}=0.08M_{\odot}$ is the minimal initial mass of a star. 

We assume that the recent specific star formation rate in our Galaxy to be a constant  \citep{belczynski+07}:
\begin{eqnarray}
\int d\bar{M}~\bar{M}\Psi(\bar{M})= 3.5M_{\odot}~{\rm yr}^{-1},\label{eq:sfr}
\end{eqnarray}
where $\Psi = \Psi_{\rm K01}\textrm{ or }\Psi_{\rm K93}$.
This assumption is justified as the life times of stars that we examine are much shorter than the evolution times of our Galaxy.

The distribution of initial binary separations is assumed to be logarithmically flat  \citep{abt83}:
\begin{eqnarray}
\Gamma(\bar{A})=\frac{\Gamma_0}{\bar{A}}.\label{eq:smadistri}
\end{eqnarray}
The normalization factor $\Gamma_0$ is determined by the range of $\bar{A}$:
\begin{eqnarray}
\int_{\bar{A}_{\rm min}}^{\bar{A}_{\rm max}}d\bar{A}\Gamma({\bar{A}})=1.
\end{eqnarray}
We set the lower limit of this integral, $\bar{A}_{\rm min}$, as the distance such that the primary fills its Roche lobe at the periastron \citep{eggleton83}:
\begin{eqnarray}
\bar{A}_{\rm min}=\frac{0.6q^{-2/3}+\ln (1+q^{-1/3})}{0.49q^{-2/3}}R_1.
\end{eqnarray}
With this definition of $A_{\rm min}$ and the normalization condition, $\Gamma_0$ is a function of $\bar{M}_1$ and $q$.
We set the upper limit as $\bar{A}_{\rm max}=10^4 $AU. Recent studies show that the double stars with the separation up to $\sim 10^4~{\rm AU}$ can stay gravitationally bound for long timescales \citep{oel17,and17}. 

As we are interested in  binaries  that contain a black hole (as a remnant of the primary star) and a secondary star that has not collapsed, the age of the system should be between $t=t_{{\rm L},1}$ and $t=t_{{\rm L},2}$, where $t_{{\rm L},i}$ is the lifetime of a star with  an initial mass $\bar M_i$.  We adopt here the lifetime suggested by  \citet{eggleton83}. For the primary star $t_{{\rm L},1}$ is the time when it collapses into a black hole.

Turning now to the current conditions of the system we note that with no current mass transfer from the secondary star to the black hole the radius of the secondary should be smaller than its Roche lobe:
\begin{eqnarray}
R_L(M_2/M_{\rm BH},A)>R_2. \label{nomasstransfer}
\end{eqnarray}
Using the mass-radius relation for terminal  age main sequence \citep{1991Ap&SS.181..313D} and the formula by \cite{eggleton83}, we can rewrite the condition (\ref{nomasstransfer}) as:
\begin{equation}
\begin{split}
A &>A_{\rm min} \\ &\equiv \frac{0.6+\left( \frac{M_2}{M_{\rm BH}}\right)^{2/3}\ln \left[ 1+\left( \frac{M_2}{M_{\rm BH}}\right)^{-1/3} \right]}{0.49} \left( \frac{M_2}{M_{\odot}} \right)^{0.83}R_{\odot}.
\end{split}
\end{equation}
This condition set the lower limit of the initial binary separation, $\bar A > \bar{A}_{\rm RL}$.  

An upper limit, $\bar{A}_{\rm period}$, is determined by the condition that the orbital period of the binary should be shorter than $P_{\rm max}$, which is given in Subsection \ref{sec:constrnts}:
\begin{eqnarray}
A<A_{\rm max}&\equiv &\left[ \frac{G(M_1+M_2)}{(P_{\rm max}/2\pi)^2}\right]^{1/3}.
\end{eqnarray}
In the following sections we related the current $A_{\rm max}$ to an upper limit $\bar A _{\rm period}$ on the initial separation.

We assume that the spacial distribution of black hole binaries in the Galaxy traces the stellar distribution.  According to \cite{1980ApJS...44...73B}, we can write the star formation number density per unit mass bin in the Galactic disk as a function of the distance from the Galactic center $r$ in the Galactic plane and the distance perpendicular to the Galactic plane $z$ as:
\begin{eqnarray}
\rho_d(r,z,\bar{M})=\Psi(\bar{M})\cdot \rho_{d,0}\exp \left[ -\frac{z}{h_z}-\frac{r-r_0}{h_r} \right],\label{eq:numdensity}
\end{eqnarray}
where $r_0=8.5~{\rm kpc}$ is the distance from the Galactic center to the Sun, and $(h_z,h_r)=(250~{\rm pc},3.5~{\rm kpc})$ are the scale lengths for the exponential stellar distributions perpendicular and parallel to the Galactic plane, respectively.  Hereafter we consider only the disk component because the binaries in the Galactic bulge would not be observed due to the interstellar absorption.  Then we can determine the normalization factor, $\rho_{d,0}$ by:
\begin{equation}
4\pi \int_0^{r_{\rm max}}rdr \int_0^{z_{\rm max}} dz~\rho_{d,0} \exp \left[-\frac{z}{h_z}-\frac{r-r_0}{h_r}\right] = 1.
\end{equation}
We describe this distribution with respect to the spherical coordinate centered at the Earth, $(D,b,l)$, where
\begin{eqnarray}
r&=&[r_0^2+D^2\cos^2 b - 2D r_0 \cos b \cos l]^{1/2}, \\
z&=&D\sin b,
\end{eqnarray}
and $D,b$, and $l$ are the distance from the Earth, Galactic latitude, and Galactic longitude. 

The total number of black hole binaries without mass accretion  detectable by \gaia\ can be obtained as a multidimensional integral over the initial primary mass, the mass ratio, the initial separation and the position
\begin{equation}
\begin{split}
N=&\frac{2 f_{\rm bin} }{1+f_{\rm bin}}
\int_{M_{\rm min,BH}}^{100M_{\odot}}d\bar{M}_1
\int_{q_{\rm min}}^{1} dq~ \Gamma_0 \Phi({q}) \\ & \times \left( t_{{\rm L},2}-t_{{\rm L},1} \right)
\int_{\rm{max}(\bar{A}_{\rm RL},\bar{A}_{\rm min},\bar{A}_{\rm det})}^{{\rm min}(\bar{A}_{\rm period},\bar{A}_{\rm max})} d\bar{A}~ \frac{1}{\bar{A}} 
\\
& \times \int_0^{2\pi}dl\int_0^{\pi/2}\cos bdb \int_0^{D_{\rm max}}D^2dD\rho_d(D,b,l,\bar{M}_1)  ,
\end{split}
\label{eq:numinted}
\end{equation}
where we set the lower limit of the integration with respect to $\bar{M}_1$ as $M_{\rm min,BH}=20M_{\odot}$, above which a primary star would form a black hole after its collapse, and $f_{\rm bin}$ is the binary fraction.  Hereafter we assume $f_{\rm bin}=0.5$, which means that we have 50 binaries and 50 single stars out of 150 stars.
In addition, $q_{\rm min}$ represents the minimum mass ratio defined in \S \ref{sec:extinction}, and $\bar{A}_{\rm det}$ is defined by considering the condition for the BH identification with \gaia\ (\S \ref{sec:constrnts}).
$D_{\rm max}$ is set as 10 kpc in \S \ref{sec:res}.
Note that while the integration looks simple, it involves numerous implicit dependences, e.g., $t_{{\rm L},i}\ (i=1,2)$ depend on $\bar{M}_i$.

As shown below, the final binary separation, $A$,  can be described as
\begin{eqnarray}
A=\bar{A}\cdot a(q),
\end{eqnarray}
where $a(q)$ is a  function of the initial mass ratio $q$, then we can describe the differential number distribution of BH binaries of interest as
\begin{equation}
\begin{split}
&\frac{dN}{dM_{\rm BH}dM_2dA dR}=\frac{dN}{d\bar{M}_1d\bar{M}_2d\bar{A}dR}\cdot \frac{\partial (\bar{M}_1,\bar{M}_2,\bar{A})}{\partial(M_{BH},M_2,A)}  \\
& \ \ =\frac{2 f_{\rm bin} \Gamma_0}{1+f_{\rm bin}}\cdot 2\int_0^{2\pi}\cos bdb\int_0^{\pi/2}dl R^2\rho_d(R,b,l,\bar{M}_1)\cdot \\ 
&\ \ \ \ \ \frac{1}{\bar{M}_1}\left( t_{{\rm L},2}-t_{{\rm L},1} \right)
  \frac{1}{\bar{A}}\cdot \frac{\partial (\bar{M}_1,\bar{M}_2,\bar{A})}{\partial(M_{BH},M_2,A)},
\end{split}
 \label{jacobian}
\end{equation} 
where $\partial (\bar{M}_1,\bar{M}_2,\bar{A})/\partial(M_{BH},M_2,A_f)$ is the Jacobian of the variable transformation from the initial to final parameters.

\subsection{Relation between the ZAMS mass and BH mass}
\label{sec:mrel}
We assume for simplicity that the BH mass satisfies the equation:
\begin{equation}
M_{\rm BH} = k \bar{M_1},\label{eq:linearrel}
\end{equation}
where we adopt $k=0.2$ as the fiducial case.
This is a conservative assumption. A good approximation is given by Equation 2.42 in \citet{egg11}. In short $k$ increases with mass from about 0.1 at $M_1=0.8\sm$ to about 0.4 
at $M_1=40\sm$, and then it flattens towards 0.5 at very large masses. In reality the WR stars will loose a lot of matter in the wind later, and the BH will have a smaller mass than the mass of $k M$. Our choice of $k=0.2$ takes in to account these two processes. 
We also try $k=0.1$ and 0.5.

However, the real relation between the ZAMS mass and BH mass should be more complicated.
Thus, we assume the mass relation as follows:
\begin{equation}
M_{\rm BH} = 
\frac{2}{\ln 3} \ln(\bar{M_1}-19)+2,
\label{eq:brokenrel}
\end{equation}
where we use the mass relation in \citet{belczynski+08} as a reference, and this function satisfies $M_{\rm BH} = 2 \textrm{ for }\bar{M_1}=20$ and $M_{\rm BH} = 10 \textrm{ for }\bar{M_1}=100$. The model with this mass relation is named ``curved''.

\subsection{Formation paths of binaries with a BH}
\label{sec:path}
A general scenario for formation of binaries with a BH involves 
a binary that initially contains a massive star and a companion.
We assume that the initial orbit is circular. The more massive star 
- the BH progenitor - will evolve faster and will initiate the first mass transfer when it sufficiently expands and fills the Roche lobe. Here, we adopt $3 \times 10^3 {\rm R}_{\odot}$ as the maximum radius of the stars with $\bar{M}_1 >20 \sm$. This value is given by the formula of the radius of the asymptotic giant branch star with mass of $20 \sm$ and the luminosity of $10^{5.5}L_{\odot}$ (Hurley et al. 2000). Although the radius depends on the mass and luminosity of a star, the uncertainty is no more than a factor of few, which has little influence on the result. We discuss this issue in detail in Section 4.
We discuss the outcome of the mass transfer in detail for two cases of the mass ratio in Sections \ref{sec:q1} and \ref{sec:qsmall}. 

In our calculation we make several simplifying assumptions. We neglect the wind mass loss from the stars.
The wind mass loss in the pre mass transfer phase will lead to tightening of the orbit, and therefore it is degenerate with the initial orbital separation. The mass loss  from the primary will also decrease the mass 
that can be transferred to the companion in the large mass ratio case.
However the contribution to the observed number of binaries in this case is small, which is discussed in detail in Section \ref{sec:discuss}.  We assume that a BH forms through direct collapse of the compact core with no mass loss during the process. Additionally,  we assume that the BH receive no natal  kicks during formation.  The current understanding of BH formation is that such kicks are small (Section \ref{sec:discuss}). 

\subsection{A large mass ratio binary evolution}
\label{sec:q1}
If the  mass ratio is larger than $\gtrsim 0.3$, the mass transfer from a giant primary star will initially be unstable but then it will stabilize because the mass ratio will be reversed.  In such a mass transfer the orbit initially tightens (i.e., the orbital separation decreases), and the system loses its mass  rapidly.  Once the mass ratio of the binary reaches unity, the separation starts to increase and the mass transfer finally ceases.  Using the formulae presented by \cite{1992ApJ...391..246P}, the ratio of  the orbital separation to the initial one is given by
\begin{eqnarray}
a(q)&=&\frac{M_{\rm BH}+M_2}{\bar{M}_1+\bar{M}_2}\left( \frac{M_{\rm BH}}{\bar{M}_1} \right)^{c_1}\left(\frac{M_2}{\bar{M}_2} \right)^{c_2},
\end{eqnarray}
where $c_1=\alpha(1-\beta)-2$ and $c_2=-\alpha(1-\beta)/\beta-2$.  Here $\alpha$ is the specific angular momentum per unit mass lost from the system, and $\beta$ is the fraction of mass that goes to the acceptor from the donor.  We assume that the mass transfer would continue until the primary mass is equal to the mass of its remnant (i.e., a BH). 
After the mass transfer, the mass of the secondary becomes $M_2=\bar{M}_2+\beta(1-k)\bar{M}_1$, and then we obtain
\begin{equation}
a(q)=\frac{(k+\beta(1-k))+q}{1+q}k^{c_1}\left( \frac{\beta(1-k)}{q}+1\right)^{c_2}.
\end{equation}
Hereafter we use $\alpha=1.0$ and $\beta=0.5$, corresponding to the standard evolution model \cite[e.g.][]{belczynski+02}.
In this case the Jacobian (Equation \ref{jacobian}) is:
\begin{equation}
\begin{split}
& \frac{\partial (\bar{M}_1,\bar{M}_2,\bar{A})}{\partial (M_{\rm BH},M_2,A_f)}=\frac{1}{k\cdot a(q)} \\
&\ \ =\frac{1+q}{(k+\beta(1-k))+q}k^{-c_1-1}\left( \frac{\beta(1-k)}{q}+1 \right)^{-c_2}.
\end{split}
\label{jacobianlq}
\end{equation}

\subsection{A small mass ratio binary evolution }
\label{sec:qsmall}
If the  mass ratio is smaller than $\lesssim 0.5$, the system will undergo a violent mass transfer from the primary to the secondary. This will leads to a common envelope (CE) phase.  During the CE phase, the binary orbital energy is used to expel the envelope.  The evolution of the binary separation can be modeled following \cite{webbink84}.  Let us assume that the primary star has a core  mass  $M_{c,1}$ and an envelope mass $M_{{\rm env},1}$, and that the initial and final orbital separations are $A_i$ and $A_f$, respectively.  From energy conservation we have
\begin{eqnarray}
\alpha_{\rm CE}\left( \frac{GM_{c,1}M_2}{2A_f}-\frac{GM_1 M_2}{2A_i} \right) =\frac{GM_1 M_{{\rm env},1}}{\lambda R_L(1/q,A_i)}, \label{eq:al_rel}
\end{eqnarray}
where $\alpha_{\rm CE}$ is the efficiency of converting the orbital energy into the kinetic energy of an envelope during the CE phase, and $\lambda$ is a parameter which is determined by the structure of the primary star.
Following  \cite{belczynski+02} and \cite{belczynski+08}, we assume that $\alpha_{\rm CE}\lambda=1$ as the fiducial case. For the massive star with 20 $\sm$ or larger, $\lambda$ can be $\sim$0.1 if the stellar radius is larger than $\sim$1 AU \citep{dom12}. Thus, the cases that $\alpha_{\rm CE}\lambda=0.1$ are also investigated in Section \ref{sec:res}. We also examine the case $\alpha_{\rm CE}\lambda=2.0$.
Then the binary separation shrinks 
by a factor of
\begin{eqnarray}
a(q) = \left[ \frac{2(1-k)}{\alpha_{\rm CE}\lambda r_L k q}+\frac{1}{k} \right] ^{-1}, \label{ceexpand}
\end{eqnarray}
where $r_L=R_L(q_1,A_i)/A_i$.  In order for the binary not to merge after the CE phase, the final separation should be larger than the sum of the stellar radii: $A_f>R^{\prime}_1+R^{\prime}_2$, where $R^{\prime}_i$ ($i=1,2$) are the stellar radii right after the CE phase.  In this case the Jacobian (Equation \ref{jacobian}) is:
\begin{eqnarray}
\frac{\partial (\bar{M}_1,\bar{M}_2,\bar{A})}{\partial (M_{\rm BH},M_2,A_f)}=\frac{1}{k\cdot a(q)} 
=\frac{1}{k}\left[ \frac{2(1-k)}{r_Lkq}+\frac{1}{k} \right].
\label{jacobiansq}
\end{eqnarray}

\subsection{Effect of the interstellar extinction}\label{sec:extinction}
\gaia\  is observing in  optical wavelengths and therefore, interstellar extinction reduces the total number of BH binaries detectable
by \gaia. 
\gaia's limiting magnitude
is 20 mag, and the average  extinction of the Milky Way disk
is $\sim$1 mag per 1 kpc in the V-band (see \citet{spi78}, and, for example,  \citet{sha17}).
Thus, the fraction of stars detectable by \gaia\  in all stars may be drastically reduced
for the distance farther from 1 kpc.
As the lower limit of the luminosity of the observable star gets higher, the lower limit of the corresponding stellar mass gets higher.
In this paper, we equate the \gaia\ band with the V-band, which is valid when
the star is bluer than G-type stars whose color V-I $\lesssim$1
\citep{jor10}.

 The interstellar extinction in the V band $A_{\rm V}$ affects the relation between the absolute magnitude $\mv$ the apparent magnitude $m_{\rm V}$, and the distance $D$:
\begin{equation}
\mv = m_{\rm V}-5(2+\log_{10}D_{\rm kpc})-A_{\rm V}(D_{\rm kpc})
\label{absmag}
\end{equation}
where the distance is normalized by 1 kpc and we adopt $A_{\rm V}(D_{\rm kpc})=D_{\rm kpc}$.
In addition, the absolute magnitude is related to the companion mass, $M_2$, as:  
\begin{equation}
 M_2 = \begin{cases} 10^{-0.1(\mv-4.8)} & \mv < 8.5 \\ 1.9 \times 10^{-0.17(\mv-4.8)} & \mv > 8.5 \end{cases},
\label{compmass}
\end{equation}
where we adopt the empirical mass-luminosity relation in \citet{smi83}. The absolute magnitude $ \mv = 8.5$ corresponds to a companion mass $M_2=0.4 \sm$. By combining Equations (\ref{absmag}) and (\ref{compmass}), we find the companion mass   whose apparent magnitude and distance are $m_{\rm V}$ and $D_{\rm kpc}$, respectively. Thus, given the limiting magnitude of \gaia\ and distance of the system, we obtain $M_{2,{\rm min}}$, which is defined to be the minimum mass of the companion observable by \gaia. Therefore, $q_{\rm min}$ in Equation (\ref{eq:numinted}) is now defined to be $\max (\bar{M_2}(M_{2,{\rm min}})/\bar{M_1}, \bar{M}_{\rm min}/\bar{M_1})$.

\subsection{Constraints required for the BH identification}\label{sec:constrnts}
 We need impose constraints on various parameters of the binaries to identify the primary of the binary system as a BH. The robust way to do so  is to measure its mass. The astrometric observations of the companion star enables us to estimate the mass of the other unseen object through the measurements of the semi-major axis and the orbital period. The mass $M_{\rm BH}$ can be expressed by $M_2$, the orbital period $P_{\rm orb}$, and 
the  angular  semi-major axis, $a_*$: 
\begin{equation}
\frac{(M_{\rm BH}+M_2)^2}{M_{\rm BH}^3} = \frac{G}{4\pi^2}\frac{P_{\rm orb}^2}{(a_*D)^3},
\label{eq:kepler}
\end{equation}
where $G$ is the gravitational constant. 
This equation means that the identification of BH requires measurements of $M_2$, $P_{\rm orb}$, $a_*$, and $D$ with a sufficient accuracy. In what follows, we estimate the required standard errors 
of these parameters for the BH identification and constraints on these parameters.

If the  mass of the hidden companion  is larger than 3 $\sm$ with a n-$\sigma$ confidence level, the object can be identified as a BH. Note that 3 $\sm$ is the fiducial minimum mass of BH expected from the maximum mass of neutron stars \citep{kal96}. This condition is expressed as
\begin{equation}
M_{\rm BH} - n \sigma_{\rm MBH} > 3 \sm,
\label{eq:bhidcond}
\end{equation}
where $\sigma_{\rm MBH}$ is a standard error of the mass estimate of the unseen primary, and we adopt $n=1$. Using Equation (\ref{eq:kepler}) $\sigma_{\rm MB}/M_{\rm BH}$ is related to $\sigma_{\rm M2}$, $\sigma_P$, $\sigma_a$, and $\sigma_D$ (the standard errors of the companion mass, orbital period, semi-major axis, and distance, respectively) as:
\begin{equation}
\label{eq:errprop}
\begin{split}
&\left( \frac{\sigma_{\rm MB}}{M_{\rm BH}} \right)^2 = \left( \frac{3}{2} - \frac{M_{\rm BH}}{M_{\rm BH}+M_2} \right)^{-2} \\
&\ \ \times \left[ \left( \frac{M_2}{M_{\rm BH}+M_2} \right)^2 \frac{\sigma_{\rm M2}^2}{M_2^2}+ \frac{\sigma_P^2}{P_{\rm orb}^2}+
\frac{9}{4}\left( \frac{\sigma_a^2}{a_*^2} + \frac{\sigma_D^2}{D^2} \right) \right],
\end{split}\end{equation}
where we assume that for all parameters the ratios of the standard errors to  the parameters themselves are smaller than 1, and the correlation between errors of these parameters can be neglected. From Equations (\ref{eq:bhidcond}) and (\ref{eq:errprop}), we can  constrain the ratios of the standard errors to the parameters, $\sigma_{\rm M2}/M_2$, $\sigma_P/P_{\rm orb}$, $\sigma_a/a_*$, and $\sigma_D/D$. If the following constraints are satisfied,  Equation (\ref{eq:bhidcond}) is also satisfied, when $M_{\rm BH} > 5 \sm$:
\begin{equation}
M_2<0.1\sigma_{\rm M2},\ P_{\rm orb}<0.1\sigma_P,\ a_*<0.1\sigma_a,\text{ and } D<0.1\sigma_D.
\label{eq:condsforparam}
\end{equation}
When $M_{\rm BH}<5 \sm$, conditions more stringent than in Equation (\ref{eq:condsforparam}) are required for satisfying Equation (\ref{eq:bhidcond}). However, empirically few BHs weigh less than 5 $\sm$ \citep{ozel+10}, so that we expect that the conditions in Equation (\ref{eq:condsforparam}) is useful for identifying most BHs. We note that these condition equations (Equation \ref{eq:condsforparam}) are given to simplify the following reduction, and therefore, just necessary conditions.

The constraints on the standard errors of orbital period, semi-major axis, and distance in Equation (\ref{eq:condsforparam}) are reduced to conditions on BH binaries detectable by \gaia.
The constraint on the standard error of the distance imposes a condition on the magnitude of the companion at a distance $D$. Since the distance is inversely proportional to the parallax $\pi$, $\sigma_D/D$ can be expressed as $\sigma_\pi/\pi$, where $\sigma_\pi$ is the standard error of the parallax for the \gaia\ astrometry, which is related to the apparent magnitude of the \gaia\ band \citep{deb12}. Assuming that the apparent magnitude of the \gaia\ band is equal to V-band magnitude as done in the previous section, the condition $\sigma_\pi/\pi$ is represented as
\begin{equation}
\left( -1.631+680.8\cdot z(m_V)+32.73\cdot z^2(m_V) \right)^{1/2} < \frac{10^2}{D_{\rm kpc}},
\label{eq:vdrelation}
\end{equation}
where the expression of $\sigma_\pi$ appears in \citet{gai16},
and the function $z(m_V)$ is
\begin{equation}
z(m_V) = 10^{0.4(\max\left[12.09, m_V\right]-15)}.
\label{eq:zmvrel}
\end{equation}
In addition, we neglect the factor including $V-I$ of the original expression of $\sigma_\pi(G)$ because this factor changes $\sigma_\pi$ only by a few percent. We note that Equation (\ref{eq:vdrelation}) gives the maximum detectable apparent $V$ magnitude $m_{V,\textrm{max}}$ for a fixed distance as
\begin{equation}
m_{V,\textrm{max}}\sim 17.5 +2.5\log_{10}\left[ \sqrt{1+\left( 0.6D_{\rm kpc}\right)^{-2}} -1 \right],
\end{equation}
where we assume that the maximum function in Equation (\ref{eq:zmvrel}) is simply equal to $m_V$, which is valid for $D_{\rm kpc} \lesssim 10$. This maximum magnitude enables us to obtain the minimum companion mass $M_{2,{\rm min}}$ by using Equations (\ref{absmag}) and (\ref{compmass}), that is, $M_{2,{\rm min}}=M_2(m_{\rm V,max})$.

The conditions for the semi-major axis and orbital period in Equation (\ref{eq:condsforparam}) constrain the range of the semi-major axis. The standard error of the semi-major axis of the stellar orbit $\sigma_a$ is expected to be similar to $\sigma_\pi$, because the semi-major axis of the stellar orbit $a_*$ is roughly the size of the orbit on the celestial sphere. 
Thus, we can assume $\sigma_a \sim \sigma_\pi$, and therefore, the constraint on the semi-major axis of the binary system $A$ can be written as
\begin{equation}
A > 10\frac{M_{\rm BH}+M_2}{M_{\rm BH}} \sigma_\pi(m_V)D \equiv A_{\rm ast}.
\label{eq:astcond}
\end{equation}
According to orbital solutions in the Hipparcos and Tycho catalog \citep{esa97}, all binaries with the orbital period less than 2/3 of the mission period of {\it Hipparcos} show the standard error of the orbital period less than 1/10 of the orbital period. Thus, we expect that for BH binaries with the orbital period less than $\sim$3 years, the orbital period will be measured with \gaia\ at the standard error less than 10\%. Therefore, we adopt 3 years as the maximum orbital period.
The minimum orbital period might be determined by \gaia's cadence for each object, which is roughly
50 days, so that we adopt 50 days as the minimum orbital period. These conditions give a range of the semi-major axis:
\begin{equation}
A(P_{\rm orb}=50\text{ days})<A<A(P_{\rm orb}=3\text{ years}).
\label{eq:pericond}
\end{equation}
The typical standard error of the stellar mass is $\sim$10\% \citep{tet11}, where stellar masses are measured by using their luminosities and temperature, so that we expect that the standard error of the stellar mass measured by \gaia is also $\sim$10\%. Therefore, $\bar{A}_{\rm det}$ is defined to be $\max (\bar{A}(A_{\rm ast}), \bar{A}(A(P_{\rm orb}=50\text{ days})))$.

In the next section, we obtain the numbers of BH binaries detectable with \gaia\ by integrating Equation (\ref{jacobian}) for various models in which the parameters are different from each other. The parameters in models are shown in Table \ref{tab:assumptions}. We show just distributions of the BH mass for models other than the fiducial one.
We have four quantities as integration variables. The integral range of $M_{\rm BH}$ are assumed to be [4 M$_{\odot}$, 30 M$_{\odot}$] in the case of the fiducial model.
The minimal companion mass $M_2$, $M_{2,\text{min}}$, is given in Subsection \ref{sec:constrnts}.  The maximal companion mass is $\bar{M_1}[1+\beta(1-k)]$ for a mass ratio larger than 0.5 or $\bar{M_1}$ for a mass ratio smaller than 0.5. 
The integral interval of the semi-major axis is the range such that Equations (\ref{eq:astcond}) and (\ref{eq:pericond}) are satisfied, where we note that the semi-major axis of all binaries should be between $\bar{A}_{\rm min}a(q)$ and $\bar{A}_{\rm max}a(q)$.

\begin{table*}
\centering
\caption{Summary of parameters in models examined in this paper. Blanks mean the same value as the fiducial model.}
\begin{tabular}{lrrrrrrrrr}
\hline \hline
Parameters/model names & fiducial & lin01 & lin05 & curved & K03 & al01 & al20 & q$-$1 & q+1 \\
\hline
Coefficient in Equation (\ref{eq:linearrel}) & 0.2 & 0.1 & 0.5 & - & & & & & \\
Index of Initial mass function &  2.3 & & & &  2.7 \\
$\alpha \lambda$ & 1.0 & & & & & 0.1 & 2.0 & &\\
Power law index of the $q$ distribution & 0 & & & & & & & -1 & 1 \\
\hline
\end{tabular}\label{tab:assumptions}
\end{table*}

\section{Results}\label{sec:res}

\begin{figure}
\centering
\includegraphics[width=8cm,clip]{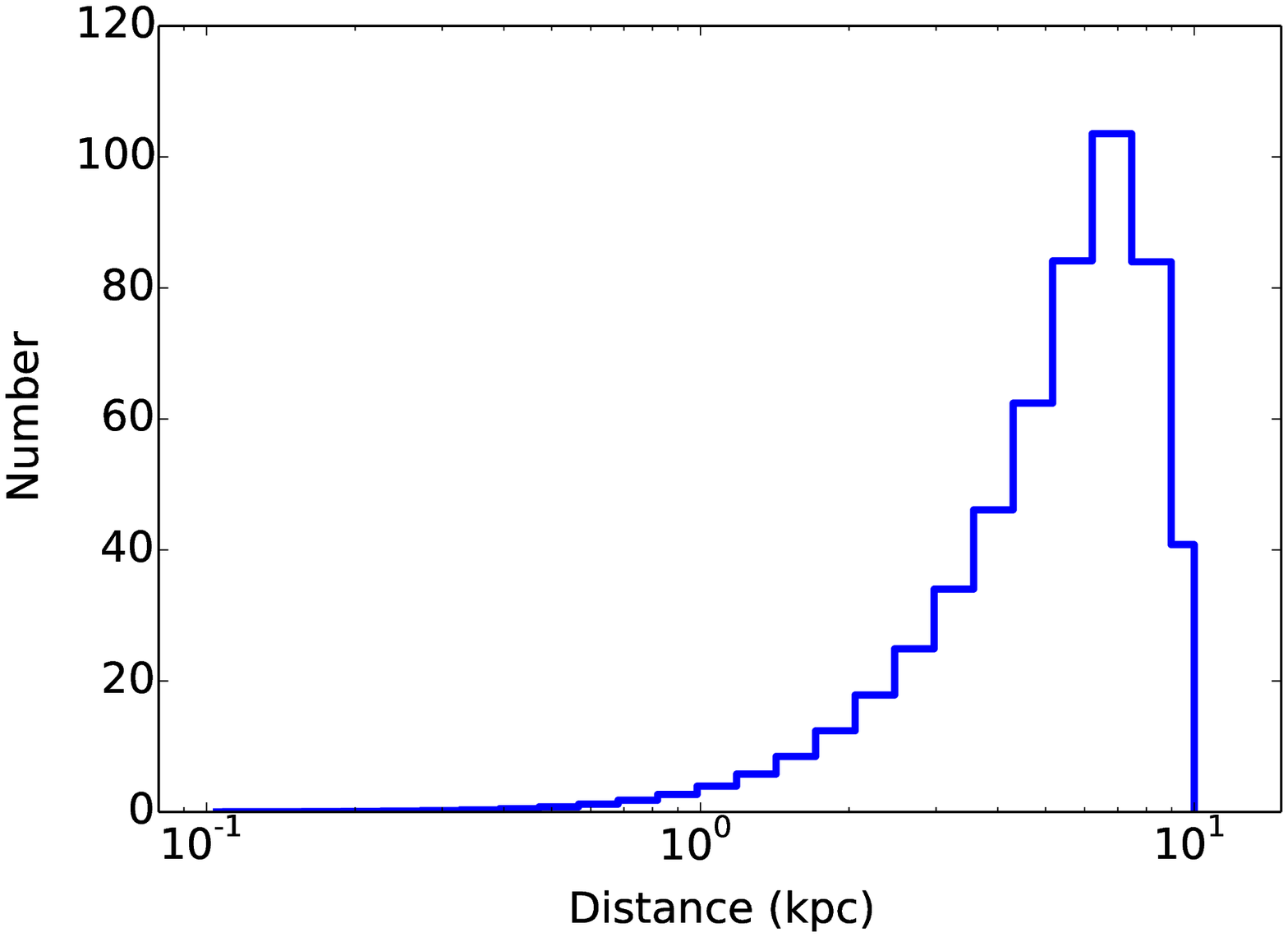}
\includegraphics[width=8cm,clip]{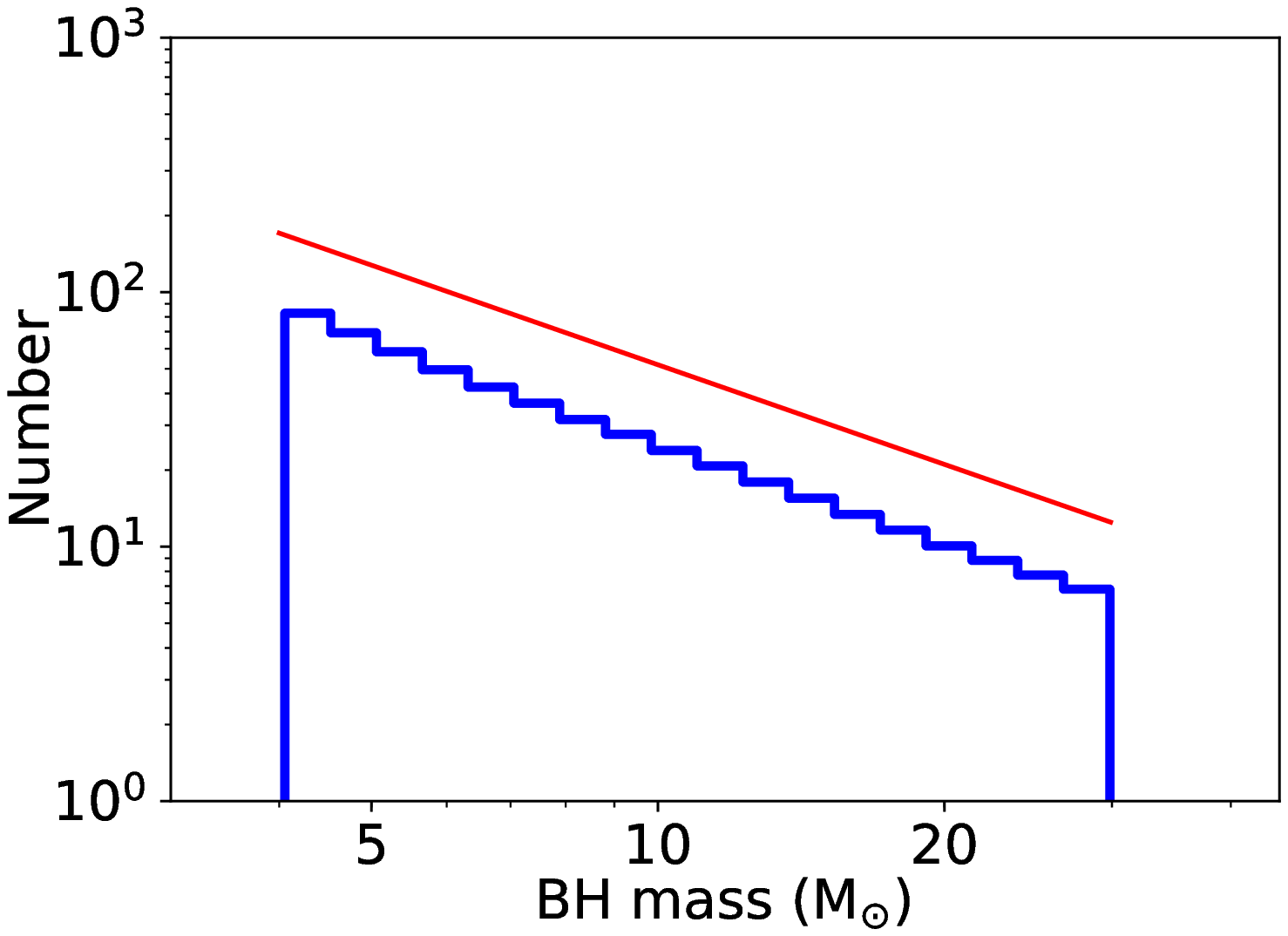}
\includegraphics[width=8cm,clip]{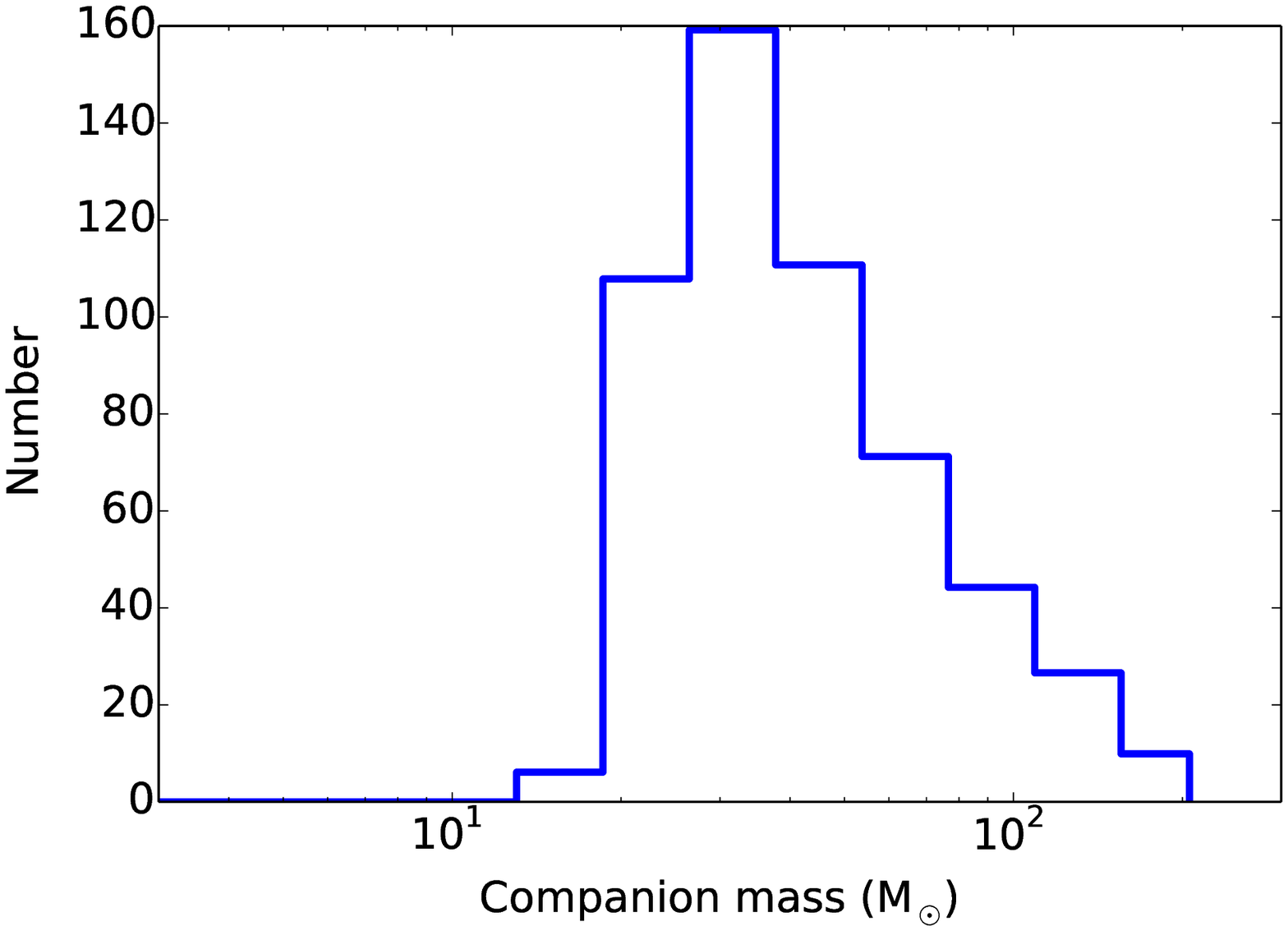}
\caption{\label{fig:fid}Calculated distributions of BH binaries in the case of the fiducial model. The upper left panel shows the distance distribution between 0.1 kpc and 10 kpc. The upper right panel and lower panel show the BH mass and the companion mass distributions, respectively. The red line in the distribution of BH mass shows the power-law function whose index is -2.3.}
\end{figure}

The calculated number of BH binaries detectable by \gaia\ for the fiducial model is $\sim$500. These include BH binaries whose companions' brightness is $m_{\rm V} \sim$20 magnitude. This number is smaller by 400 than the one found by \citet{mashianloeb17} and by about one order of magnitude  than the one found by \citet{2017ApJ...850L..13B}. In our calculation, $\sim $1 million BH-main sequence binaries exists in whole Milky Way. This is easily estimated from the star formation rate and initial mass function. The companion stars in a few percent of those are observable with \gaia, i.e., the companions are brighter than 20 magnitude for G-band. Furthermore  \gaia\ can detect the orbital motion of a few percent of those and identify them as BH binaries.

Figure \ref{fig:fid} depicts the distributions of BH binaries that are detectable by \gaia.
Almost all detectable BH binaries are within 1-10 kpc. The peak is at $\sim$7 kpc. The distribution within $\sim$5 kpc increases monotonically, corresponding to the increasing volume.  Above $\sim$7 kpc the number of BH binaries drastically decreases, because \gaia\ cannot measure the distance and semi-major axis of most binaries accurately enough to identify those objects as BH binaries. 

The upper right panel shows the power-law distribution of the BH masses. The index of this power-law distribution is $\sim -$2.3, which is just the same as that of IMF.
%
The distribution of companions masses is shown in the lower panel in Figure \ref{fig:fid}. The contribution of companions less massive than 20 $\sm$ is much smaller than that of those larger than 20 $\sm$. Binaries  with $q< 0.3$ undergo a CE phase in which its separation
decreases typically down to 1\% of the original one (see Equation (\ref{ceexpand}). Therefore,  these binaries won't  be detected by \gaia\ because of their short orbital period (or small separation). This result is different from one obtained by \citet{2017ApJ...850L..13B}. Although the minimal companions mass in the binaries that undergo a mass transfer phase is 7 $\sm$, after this phase  the companions receives a fraction of the matter of the primary.  This leads to a minimal companion mass of 15 $\sm$ used in our fiducial model. 
We also see that the maximal mass of the companions reaches $\sim 200 \sm$, although the maximal mass of primaries is 150 $\sm$. This is because when the mass ratio is larger than 0.3, we assume that the companion mass increases due to the mass transfer (see Section \ref{sec:q1}).

\subsection{Dependence of the distribution of BH mass on the models}
The total numbers of detectable BH binaries for the models, lin01, lin05, and curved, are shown in Table \ref{tab:numbers}. We see that the total number correlates with the coefficient $k$. This can be interpreted to be an effect of astrometric observation. If BH mass is larger, the orbit of companion is also larger, which increases the detectability of BH by \gaia. This can be also understood by Equation (\ref{eq:astcond}). Thus, as the coefficient $k$ increases, the total number increases.

\begin{figure}
\centering
\includegraphics[width=8cm,clip]{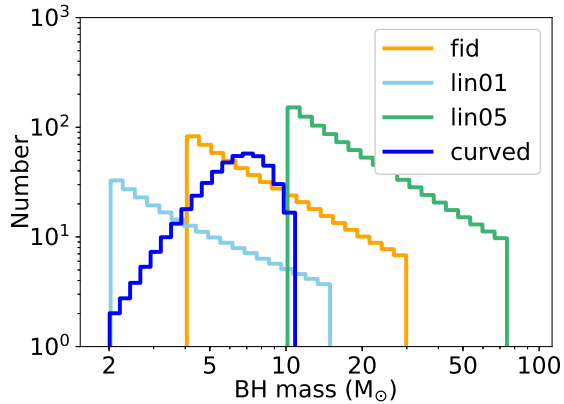}
\caption{\label{fig:bhmass_massrel}
Distributions of BH mass for different ZAMS mass-BH mass relations. We show the distribution for models of fiducial (orange), lin01 (skyblue), lin05 (green), and curved (blue).  
}
\end{figure}

Figure \ref{fig:bhmass_massrel} shows distributions of BH mass when the relation between the ZAMS mass and BH mass is changed. In the model of the linear mass relation (fiducial, lin01, and lin05), the BH masses show the single power-law distributions, and their power-law indexes are similar to each other. On the other hand, the distribution of the model ``curved'' shows a peak at $\sim 7 \sm$.
This feature arises clearly due to the projection of IMF onto the BH mass with the relation of Equation (\ref{eq:brokenrel}). The small number below 7 $\sm$ reflects the steep slope in the function of Equation (\ref{eq:brokenrel}), where we note that the minimal BH mass is 2 $\sm$, which is within the mass range of a neutron star.
The drastic decrease over 7 $\sm$ is caused by the projection of IMF whose mass range is [40 $\sm$: 150 $\sm$] onto the narrow mass range, i.e., [7 $\sm$: 10 $\sm$].

\begin{figure*}
\centering
\includegraphics[width=8cm,clip]{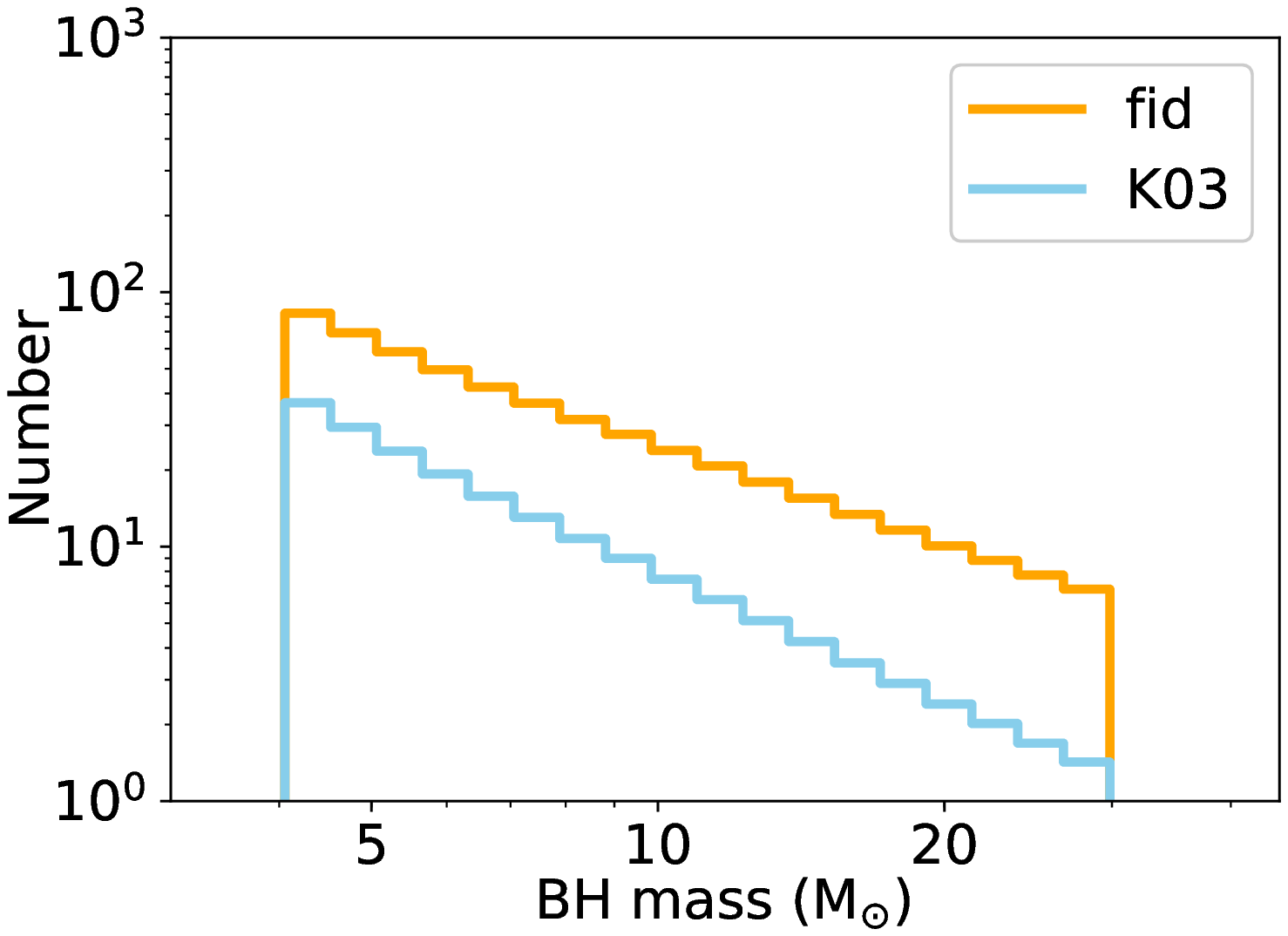}
\includegraphics[width=8cm,clip]{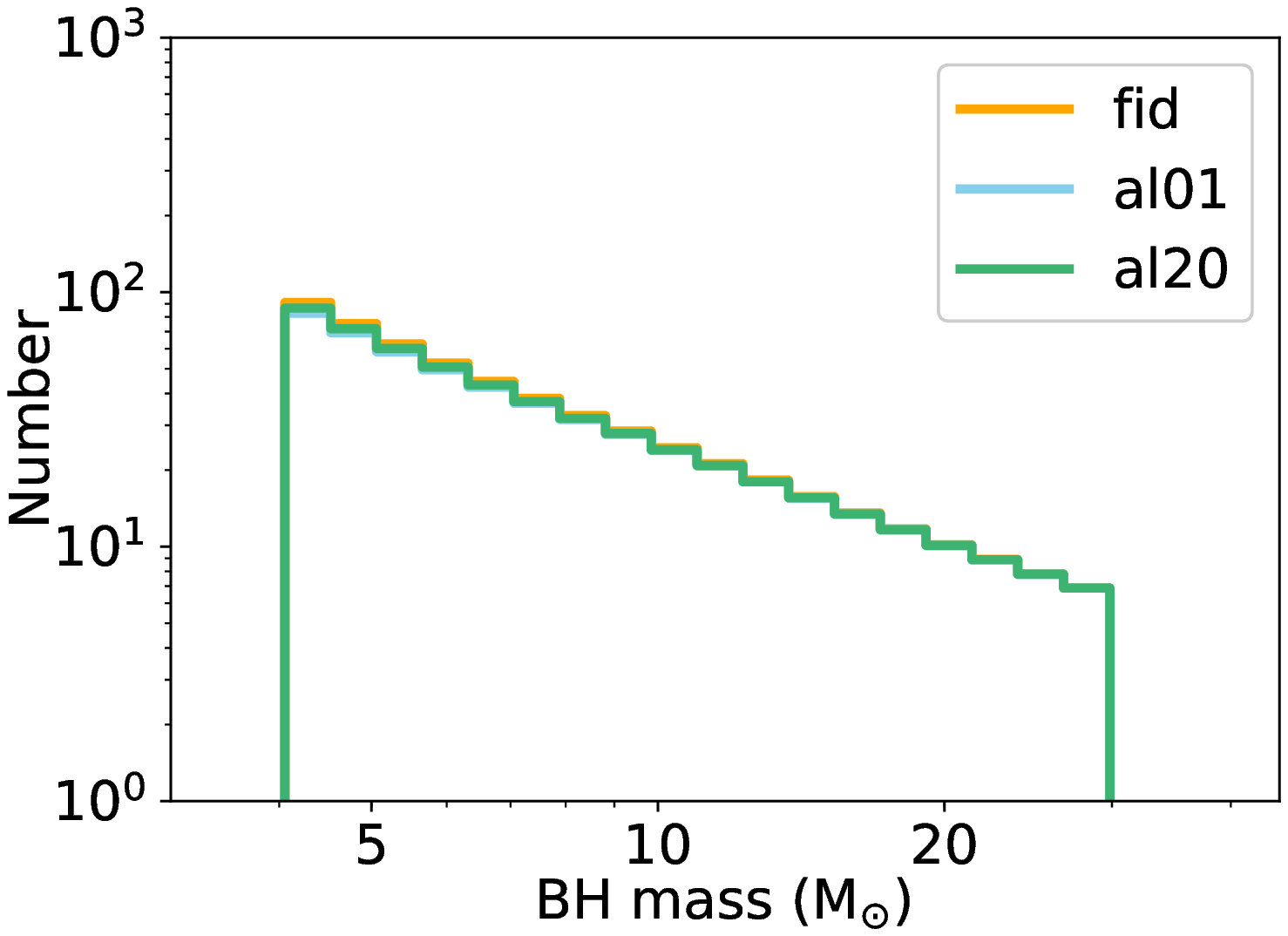}
\caption{\label{fig:bhmass_imf-al}Distributions of BH mass for fiducial (orange lines), K03 (skyblue line in the left panel), al01 (skyblue line in the right panel), and al20 (green line in the right panel). }
\end{figure*}

We also examine the  IMF shown in \citet{kroupaweidner03}. The total number of detectable BH binaries is estimated to be $\sim$200 (Table \ref{tab:numbers}). The power-law index of this IMF for stars more massive than  $1.0 \sm$ is smaller than that of the fiducial one, so that the total number of primary star  whose main sequence  mass is $>20 \sm$ is less than that of the fiducial case. This leads to a smaller number of the detectable BH binaries than in the fiducial case. This smaller power-law index also affects the distribution of the BH mass, which is shown in the left panel of Figure \ref{fig:bhmass_imf-al}. The slope of the distribution in the case of K03 is steeper than that of the fiducial case. Thus,  the power-law index of IMF directly affects the  mass distribution of BH detectable with \gaia.

The parameter $\alpha \lambda$ does not affect the total number of detectable BH binaries. This is because the detectable binaries do not undergo the CE phase. Therefore,  the distributions of BH mass (Figure \ref{fig:bhmass_imf-al}) show no difference between fiducial, al01, and al20 cases.

\begin{figure}
\centering
\includegraphics[width=8cm,clip]{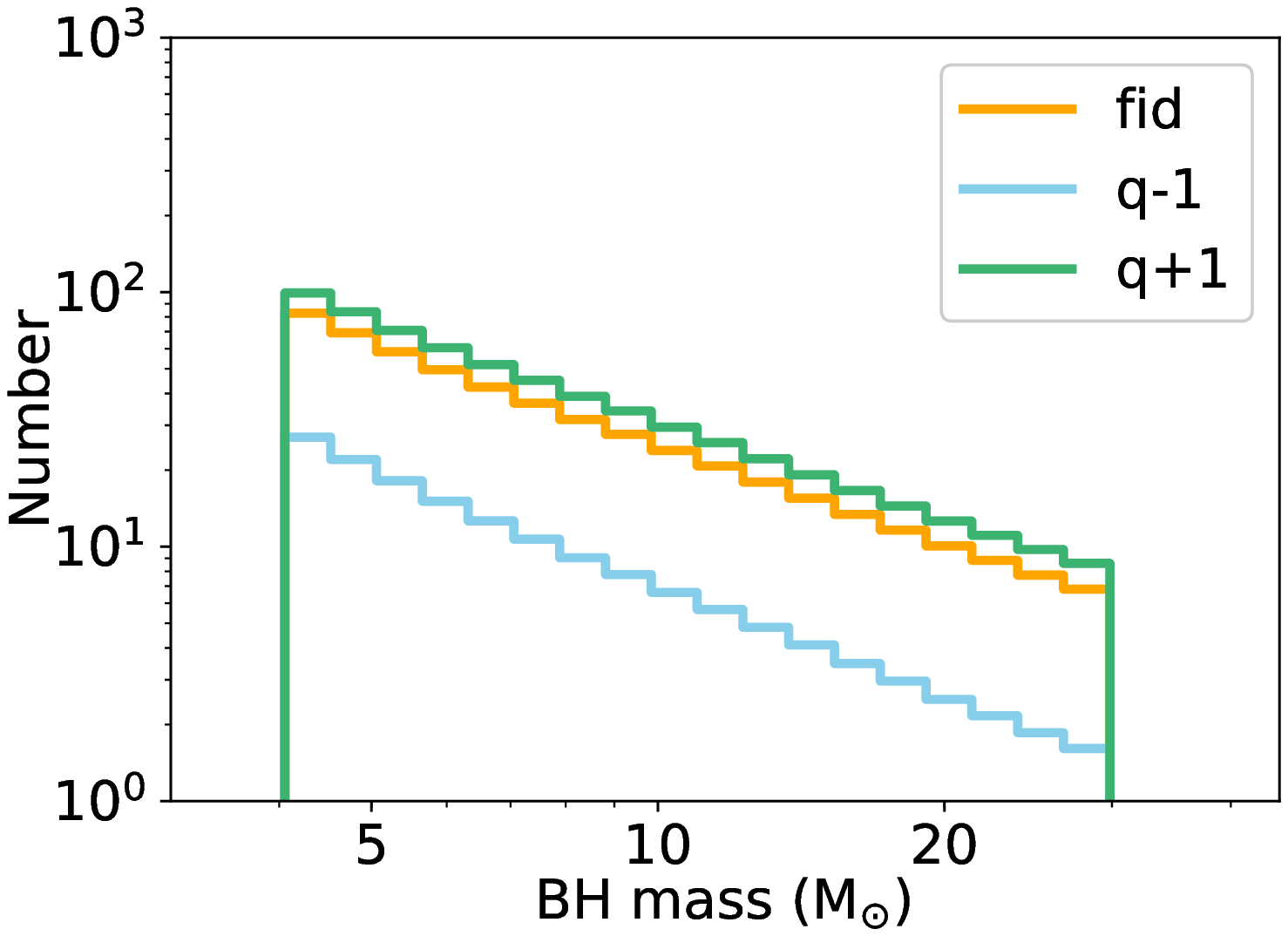}
\caption{\label{fig:bhmass_q}Distribution of BH mass for the models, fiducial (orange), q$-$1 (skyblue), and q+1 (green).}
\end{figure}

Table \ref{tab:numbers} also shows the total numbers of the cases in which the distribution of mass ratio $q$ is different. This indicates that the smaller the power-law index of the $q$ distribution, the smaller the total number. As the index becomes smaller, the fraction of massive stars becomes smaller. The massive stars contribute the total number because they undergo the mass transfer phase, so that the total number decreases as a result. Figure \ref{fig:bhmass_q} shows the corresponding distributions of BH mass, whose slopes are not so different from that of the fiducial case. This means that the $q$ distribution does not affect the shape of the distribution of BH masses.

\begin{table}
\centering
\caption{Total number of BH binaries detectable with \gaia\ for each model, which is rounded to the nearest hundred.}
\begin{tabular}{lr}
\hline \hline
Models & Total number\\
\hline
fiducial & 500 \\
lin01 & 200 \\
lin05 & 1000 \\
curved & 500 \\
K03 & 200 \\
al01 & 500 \\
al20 & 500 \\
q$-$1 & 200 \\
q+1 & 700 \\
\hline
\end{tabular}\label{tab:numbers}
\end{table}

\section{Discussion}\label{sec:discuss}

The obtained result show that the BH mass distribution function that would be obtained by \gaia\ strongly depends on the relation between the mass of a black hole and the ZAMS mass of its progenitor.  When we change the IMF and $q$ distribution, the difference in the distributions of BH masses is just the power-law index and/or the total numer. In the case of $\alpha \lambda$, the BH mass distribution is nearly unchanged. On the other hand, when we adopt the curved function as the relation between BH mass and ZAMS mass, the shape of the distribution function of BH mass has been drastically changed. This means that we can determine whether the mass relation is a linear function or a curved function as shown in \citet{belczynski+08}. Of course, we have not yet examined the entire parameter space, so that another parameter change, such as a curved IMF or more complicated $q$ distribution, can produce the distribution of BH mass with a peak. Nevertheless, IMF estimated from observations of massive stars in local universe shows single power-law distribution in $\sim 10-100 \sm$ \citep[e.g.,][]{gar82,hum84}, and the results in the last section shows weak dependence on $q$ distribution, so that we argue that from the BH mass distribution, we can constrain the mass relation.

\citet{2017ApJ...850L..13B} find different results from ours. Their histogram of the companion mass indicates that the binaries whose companion mass is less than 10 $\sm$ are included, which is because in their calculation the binaries undergoing the CE phase are included in the detectable BH binaries. In addition, the distribution of BH mass shows a complicated shape, but it seems to be consistent with the BH mass distribution of our fiducial model.

\subsection{Validity of our model}
The number estimate of detectable BHs  depends on various parameters other than those changed in this paper. Among those are  the star formation rate (Equation \ref{eq:sfr}), the distribution of semi-major axis (Equation \ref{eq:smadistri}), the binaries spatial distribution  (Equation \ref{eq:numdensity}), the range of the orbital period required for the detection of BHs (Equation \ref{eq:pericond}), and \gaia's astrometric precision $\sigma_\pi$. Thus, the total number shown in Table \ref{tab:numbers} can  easily vary if those value are different from the one that we used. 
However we do not expect that reasonable variation of these  parameters would change significantly the number of detected BHs.  

Our calculations rely on several simplifying assumptions: we ignored the effects of the stellar wind mass loss, natal kicks, and the initial eccentricities.  Before concluding we review these assumptions here. First we note that   the orbital expansion due to the stellar wind mass loss does not affect the number count of detectable BH binaries significantly.  The factor by which the orbital separation expands depends only on the ratio of the binary mass before mass loss to that after the mass loss, and it does not depend on the separation.  Since we assume a logarithmically flat distribution of initial separation, such a modification of the orbital separation does not change the final count of detectable BH binaries.  Second, the stellar wind mass loss may affect whether the binaries would experience a CE phase or a MT phase because the mass ratio would be modified from that in the ZAMS phase.  However, when the mass is maximally lost from the primary star by the time of the CE or mass transfer (MT) phase, the mass would be $k\bar{M}_1$, and then the initial mass ratio that divides two cases (CE or MT) would be no less than $\gtrsim 0.3k$.  In the fiducial model the distribution of the initial mass ratio is flat, and then such a modification would affect the final result only by a factor.  Third, as for the effect of natal kicks, Fig. 4 of \cite{2017ApJ...850L..13B} shows that kicks do not affect the final results such as the number of detectable BH binaries or its mass distribution.  Finally, as for the initial eccentricity, Fig. 4 of \cite{2017ApJ...850L..13B} shows that most of the detectable BH binaries have eccentricities that are nearly equal to zero. 
This means that our estimates obtained by assuming circular orbits for all binaries do not deviate so much from the estimates obtained  taking into account the non-zero eccentricity.

While we assume the uniform binary fraction (50\%) for entire stellar mass for simplicity,  recent studies show that massive stars are expected to be found in binaries at higher rates \citep[e.g.,][]{san12,moe17}. A larger   binary fraction  increases linearly the number of BH binaries. If we assume the binary fraction $\sim 0.7$ \citep{san12}, we can easily obtain the total numbers of BH binaries by multiplying those in Section \ref{sec:res} by $\sim$ 1.2.

We adopt a uniform value of $3 \times 10^3 {\rm R}_{\odot}$ as the maximal radius of massive stars. Although this value depends on the stellar mass (and the luminosity) and can be larger for more massive stars, a variation of this value does not affect the results. Our results show that the detectable binaries should have passed through a MT phase, which implies that the separation does not change so much after that phase for a binary with a massive companion. In addition, the orbital period of the binary whose Roche radius is $\sim 3 \times 10^3 {\rm R}_{\odot}$ is $\sim$10 yrs. This is larger than the period of the detectable binaries, which is given in Section \ref{sec:constrnts}. Thus, even if the maximal radius of massive stars is larger than $3 \times 10^3 {\rm R}_{\odot}$, \gaia\ cannot identify them as BH binaries, and therefore the results are not influenced by this uncertainty.

We assume that the G-band magnitude is equal to the V-band magnitude. This  is basically valid for nearby blue stars. Our results  show that the companions of all detectable binaries are blue stars, but most of them are located at $\sim 7$ kpc, which means that these stars suffer from significant interstellar extinction. The extinction can be estimated to be $A_{\rm V} \sim 7$ using ``1 magnitude per kpc''. The ratio of extinction $A_{\rm G}/A_{\rm V}$ given in \citet{jor10} shows that $A_{\rm G}/A_{\rm V} \sim 0.9$ for the bluest stars ($V-I \sim$0.4; $T_{\rm eff}=50000$K in \citet{jor10}) if $A_{\rm V}=5$.  Correspondingly we expect that, due to the linearity of $A_{\rm G}/A_{\rm V}$ with respect to $A_{\rm V}$,  $A_{\rm G}/A_{\rm V}\sim 0.8$ for  $A_{\rm V}\sim$ 7. Thus, we expect $A_{\rm G} \sim$ 6. This difference in  the extinction does not affect the results because the apparent V magnitude of these stars is $\sim$ 13 ($L_{\rm star}\sim 10^5$ and distance $\sim$ 7 kpc), which is much brighter than \gaia's limiting magnitude.

\subsection{Note on astrometric measurements}
The semi-major axis and the parallax are not generally degenerate in the \gaia\ astrometric measurements.
There are two reasons. One is that the periods of projected motion are different. The period of parallax is just 1 year and  the orbital period is generally different from that. Thus, we can separate these two motions by the Fourier analysis.
The other is that the phase and shape of these two motions are different. The phase and shape (ellipticity and position angle) of motion due to parallax is determined by the ecliptic longitude and latitude, respectively. On the other hand, those of the orbital motion are generally independent of the ecliptic coordinate. In this paper, we assume that the astrometric signature due to the orbital motion is large enough to be detected, so that these two motions can be separated.
Of course, we note that if the orbital period is just the same as 1 year and if the shape of orbital motion is the same as that of parallax, which occurs when the orbit is circular (the same as Earth orbit) and the inclination angle is equal to the ecliptic latitude, we expect that these two motions are degenerate. However, such case must be rare.

\section{Conclusion}\label{sec:conclusion}

We have investigated the prospect of detecting of Galactic black hole binaries by \gaia by observing the orbital motion of the BH's companion.  We have taken into account the orbital change due to a mass transfer and a common envelope phase. In addition,  we have taken into account  interstellar absorption that was not considered before. We have calculated the distribution of  BH binaries detected by \gaia\ adopting a signal to noise ratio of 10 for both the semi-major axis and the parallax.  We show that, assuming our fiducial model,  \gaia\ will be able to identify $\simeq$ 500 stars as companions of black holes. varying various parameters in this model we find the uncertainty of this estimate is in the range $200-1,000$. This values are much smaller than   those obtained by \citet{mashianloeb17} and by \citet{2017ApJ...850L..13B}. 

The shape of the mass distribution function of the detectable black holes depends strongly on the relation between the ZAMS mass of a star $\bar{M}_1$ and its remnant mass $M_{\rm BH}$. This mass relation is difficult to estimate from observations because the number of identified Galactic black hole binaries is still small and their ZAMS is not known.  Our results shows that the black hole mass distribution obtained by \gaia will enable us to   estimate this  ZAMS mass-black hole mass relation.   This relation is important for the understanding of the stellar evolution process as well as understanding the core collapse process in  massive stars.

The research was supported by JSPS KAKENHI Grant Number JP15H02075 and JP18K13576, the HAKUBI project at Kyoto University, the I-Core center of excellence for Astrophysics, and an advanced ERC grant TReX. TB acknowledges the NCN grant UMO-2014/14/M/ST9/00707.

\end{document}